# Insights of preferred growth, elemental and morphological properties of $BN/SnO_2$ composite for photocatalytic applications towards organic pollutants

Bikramjeet Singh[1], Kulwinder Singh[1], Manjeet Kumar[2], Akshay Kumar[*1]

[1]*Advanced Functional Materials Lab. Department of Nanotechnology, Sri Guru Granth Sahib World University, Fatehgarh Sahib-140 407, Punjab, India*

[2]*Department of Electrical Engineering, Incheon National University, Incheon, 406772, South Korea*

**Email**: akshaykumar.tiet@gmail.com

## Abstract

Boron nitride (BN) has been explored these days because of its extraordinary optical, chemical and mechanical properties. BN is sensitive to its crystal structure that slight change in lattice parameters enormously change its properties. Present study deals with synthesis, characterization as well as photocatalytic applications of BN-based composite. When boron nitride was mixed with $SnO_2$ having tetragonal crystal structure, dissociation into smaller sheets occurred and the material oriented to (102) plane. $SnO_2$ particles attached both sides of BN sheets provided high surface area which make the material suitable for catalytic process. Presence of large number of active sites leads to the formation of hydroxyl radicals in $BN/SnO_2$ composite which helps during degradation of organic and colourless pollutants i.e. methyl orange dye up-to ~92% under 7 minutes and salicylic acid in 40 minutes to ~82%. Results suggested that $BN/SnO_2$ composite material possesses good capability for use in environmental as well as industrial applications.

## 1. Introduction

Photocatalytic technology signifies an easy way to utilize the energy of solar light. It has been extensively used in solar energy transformation, environmental shield and is an auspicious, environmental, cost-effective method [1-3]. As the environmental pollution has become a major challenge to the progress of modern human society so the waste water treatment utilizing eco-friendly photocatalysis has been broadly studied. In photocatalytic process the electron-hole pairs are generated by means of band gap radiation that can give rise to redox reactions to the species adsorbed on the surface of the photocatalyst. By reduction of particle size and modifications in surface structure, the photocatalytic activity of semiconductor materials can be considerably enhanced [4-6]. To increase the catalytic activity, non-oxide catalyst support has been used in place of oxide supported metal catalysts [7,8]. The major limitation is that most of these catalysts that they are active only under UV irradiation. Some are absorbing visible light but are not stable during the reaction process, such as photo corrosion [9]. Based on first principle calculations, semi-hydrogenated BN sheet (Sh-BN) with a band gap of 2.24 eV has been reported as another candidate of visible-light driven photocatalysts which is simple, reachable and containing no transition metals [10]. As a result, h-BN has a wide range of applications, such as deep ultraviolet light emitters, transparent membranes, hydrogen storage and organic/inorganic pollutant adsorption catalysis [11]. After introduction of h-BN, the photocatalytic performance of the bulk semiconductors has been improved successfully [12] Because of the low density, excellent chemical inertness and environmental friendliness of boron nitride-based catalysts, several reports have been published about good catalytic applications, including ammonia synthesis, deep oxidation of volatile organic compounds, $NO_x$ reduction, selective hydrogenation, hydrogen production [13-16] etc.

Layered structure of h-BN is beneficial in reducing the rapid recombination of photo generated electron-hole pairs from photocatalyst i.e. necessarily condition for photocatalytic activity [17]. It is advantageously to use h-BN as a photocatalyst support. Presence of the h-BN promoted the separation of electrons and holes in the photocatalytic reaction. Therefore, h-BN can act as a supporting matrix and a metal-free co-material [18]. Tang et al. improved the photocatalytic reduction of $TiO_2$ by using the intrinsic electrostatic potential of BN nanotubes [19]. Meng et al. reported the enhanced photocatalytic activity of BN/metal sulfide towards aromatic alcohols to aromatic aldehydes [20]. In addition, Chen et al. successfully synthesized h-BN/$TiO_2$ and h-BN/ZnO composites under ball milled condition that displayed good photocatalytic activities towards degradation of organic dyes [13, 21]. $SnO_2$ possess larger band gap ($E_g$= 3.6) and have shown photocatalytic properties towards organic dyes [22,23]. But high electron-hole recombination rate reduces the efficiency of photocatalysis process. To overcome this limitation of $SnO_2$ photocatalyst, current studies focus on the immobilization of the $SnO_2$ onto suitable supports. Many efforts have been made to improve the performances of nanostructured $SnO_2$. Wang et. al [24] reported a decent photocatalytic property of BN/$SnO_2$ materials was principally due to its solid adsorption capability for methyl orange, appropriate band gap energy and compelling charge partition at the BN/$SnO_2$ photocatalyst interface. By making composite with BN, it can either change absorption edge or acts as a carrier for transferred hole leading to suppression of photo generated electrons/holes and consequently improves the photocatalytic activity.

In this paper, BN/$SnO_2$ composites have been successfully synthesized by a hydrothermal cum-wet chemical method. Structural, morphological, elemental and photocatalytic properties of synthesized material have been studied. Results showed that the material is preferentially oriented toward (102) plane which provides high surface area, consequently large number of surface-active sites are present. X-ray photoelectron spectroscopy (XPS) analysis confirmed the presence of different elements in the composite. BN/$SnO_2$ composite shows highest photocatalytic activity compared with common photocatalysts i.e. BN and $SnO_2$. Results showed that BN/$SnO_2$ composite almost degraded methyl orange (MO) dye within seven minutes under visible light irradiation.

## 2. Experimental Details

### *2.1 Materials and methods*

All the chemicals used for synthesis process were purchased from Sigma Aldrich and used as obtained. Hydrothermal method [25, 26] was employed for the synthesis of BN nanostructures. Full description of BN synthesis procedure can be found in our previous report [26]. In a typical BN hydrothermal synthesis process, boric acid ($H_3BO_3$) and ammonia solution ($NH_4OH$) were mixed and put into a stainless-steel autoclave. Autoclave was sealed and heated from room temperature to 700 ˚C for 24 h. After the autoclave was cooled down to room temperature, the product was collected and washed. Finally, the product was dried in a vacuum oven at 50 ˚C for 5 h.

To synthesize BN/$SnO_2$ composite, common wet chemistry method was employed [24, 27]. Tin chloride ($SnCl_2$) powders were dissolved in distilled water and then HCl and BN nanostructured were added respectively. The mixed solution was magnetically stirred for 3 h at 100 ˚C. Then, the reaction mixture was allowed to cool to room temperature. To collect the resultant precipitates, the mixture was repeatedly washed with ethanol. The precipitates collected were dried at 70 ˚C for 6 hr in oven to get the resultant powder.

### 2.2 *Characterization*

Synthesized BN/$SnO_2$ composite was characterized by powder X-ray diffraction (XRD). XRD patterns were recorded in a Bruker instrument with Cu $K_\alpha$ radiation ($\lambda$ = 1.5418 Å) from 10˚ to 80˚ at a slow scanning rate. Field Emission Scanning Electron Microscope (FESEM) was carried out on Hitachi at 15.0 kV. Elemental studies were determined using X-ray photoelectron spectroscopy (XPS). Quantachrome Novae-2200 was used for the determination of surface area of synthesized sample through nitrogen adsorption–desorption isotherms. Optical absorption spectrum was studied using UV-Visible Shimadzu UV-2600 instrument. Photoluminescent measurements were performed using photo-luminescence (PL) spectrometer (Edinburgh FLS-890).

2.3 *Catalytic experiment set up*

The photocatalytic activities of $BN/SnO_2$ were evaluated by degradation of methyl orange (MO) water soluble dye and colourless pollutant i.e. salicylic acid. All experiments were carried out at room temperature. Absorbance of the suspension was studied using Shimadzu UV-2600 spectrophotometer. The maximum absorption wavelength of MO was observed at 464 nm. Typically, 25 mg /L of photocatalysts ($BN/SnO_2$) was added into 50 mL of 10 mg/L MO aqueous solution. Firstly, the suspension was stirred under dark condition for 30 min to establish an adsorption-desorption equilibrium. Samples were taken out after particular time and centrifuged to separate photocatalyst for absorption readings. Degradation rate [28, 29] was calculated as:

$$D = \frac{A_0 - A}{A_0} \quad (1)$$

Here '$A_0$' is initial absorbance and 'A' shows sample absorbance. All the experiments were carried out three times and used for calculation of error limits which has been shown in the figures as error bar.

2.4 *Analysis of Hydroxyl Radicals ($OH^{\bullet}$)*

Hydroxyl radicals ($OH^{\bullet}$) produced during the photocatalysis under visible light irradiation was estimated by the PL spectroscopy using terephthalic acid (TA) as a probe molecule. In a typical process [30], 10 mg of catalyst was dispersed in 30 mL of $5 \times 10^{-4}$ M of TA and diluted aqueous

NaOH ($2 \times 10^{-3}$ M) solution. The resulting suspension was then exposed to visible light irradiation. At regular intervals, 2.5 mL of the suspension was collected and centrifuged to measure the maximum PL emission intensity with an excitation wavelength of 315 nm. This method relies on the luminescence signal at 425 nm of 2-hydroxyterethalic acid (TAOH).

## 3. Results and discussion

XRD pattern of BN is reported in our previous article [27]. In general, the reflection peaks at 27 °, 42 ° and 51 ° were indexed to planes (002), (100) and (102) of hexagonal BN, the calculated lattice constants (a = 2.41 Å and c = 6.06 Å) are in close proximity with ICDD card no. 45-0896. XRD analysis showed that the reflection peaks are shifted slightly to a large diffraction angle compared to the standard data card which indicated the lattice compression between two adjacent BN slabs along the c axis. Fig. 1 shows the XRD pattern of the BN/$SnO_2$ composite and the diffraction peaks have been assigned to $SnO_2$ and BN with reference ICDD card no. 41-1445 and 45-0896 respectively. The broadening of diffraction peaks as well as weak intensity was occurred, indicated the high dispersion and small dimensions of $SnO_2$ and BN particles [24]. Crystallite size of BN/$SnO_2$ composite was found to be ~15.3 nm (calculated using Debye-Scherrer formula).

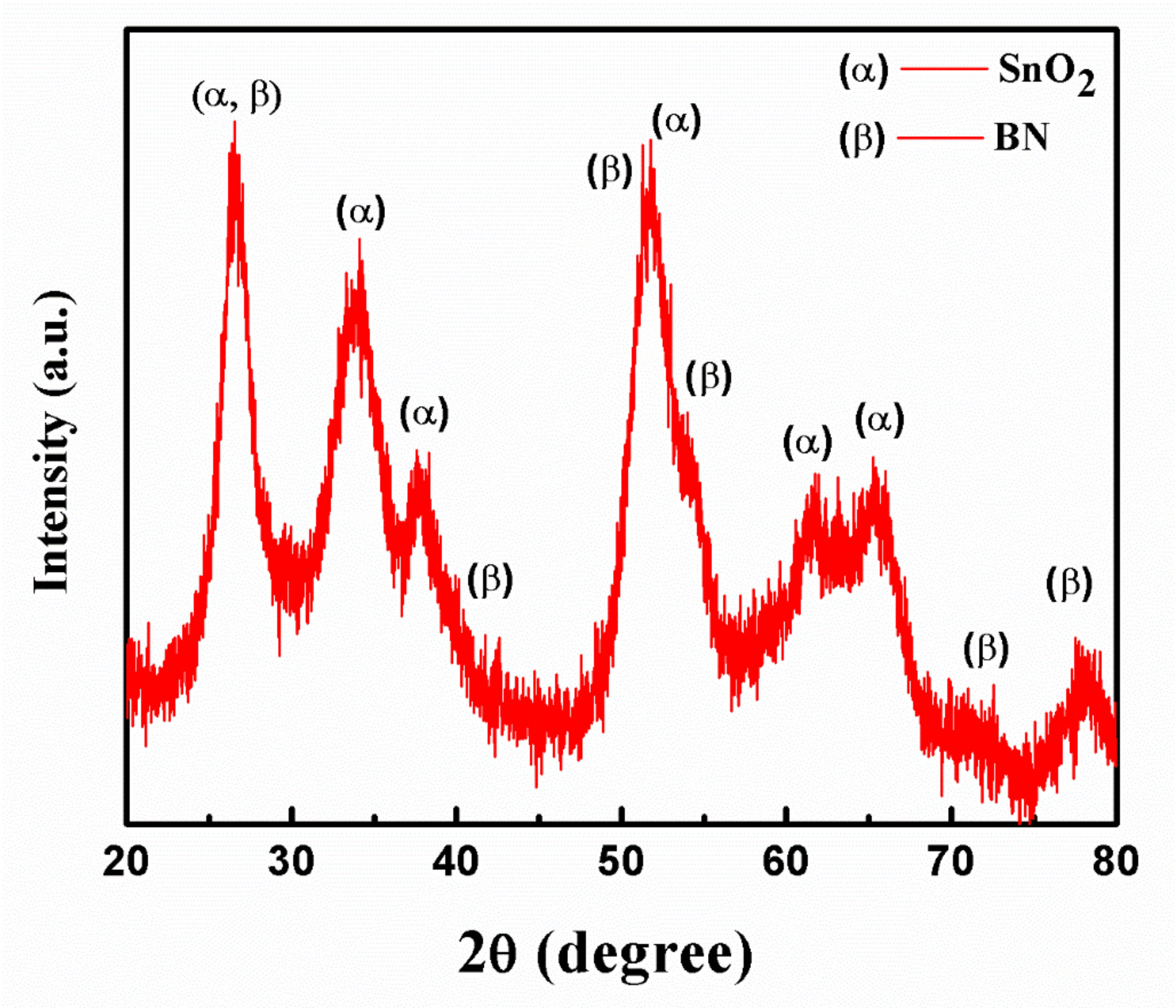

Fig. 1 XRD pattern of composite $SnO_2$/ BN

XRD analysis revealed that the particle size is small in case of BN/$SnO_2$ composite. BET analysis showed that the BN/$SnO_2$ composite possesses high surface area (45.64 $m^2/g$) shown in Fig. 2. Barrett–Joyner–Halenda method was applied for determining the pore size i.e. diameter ~35 Å.

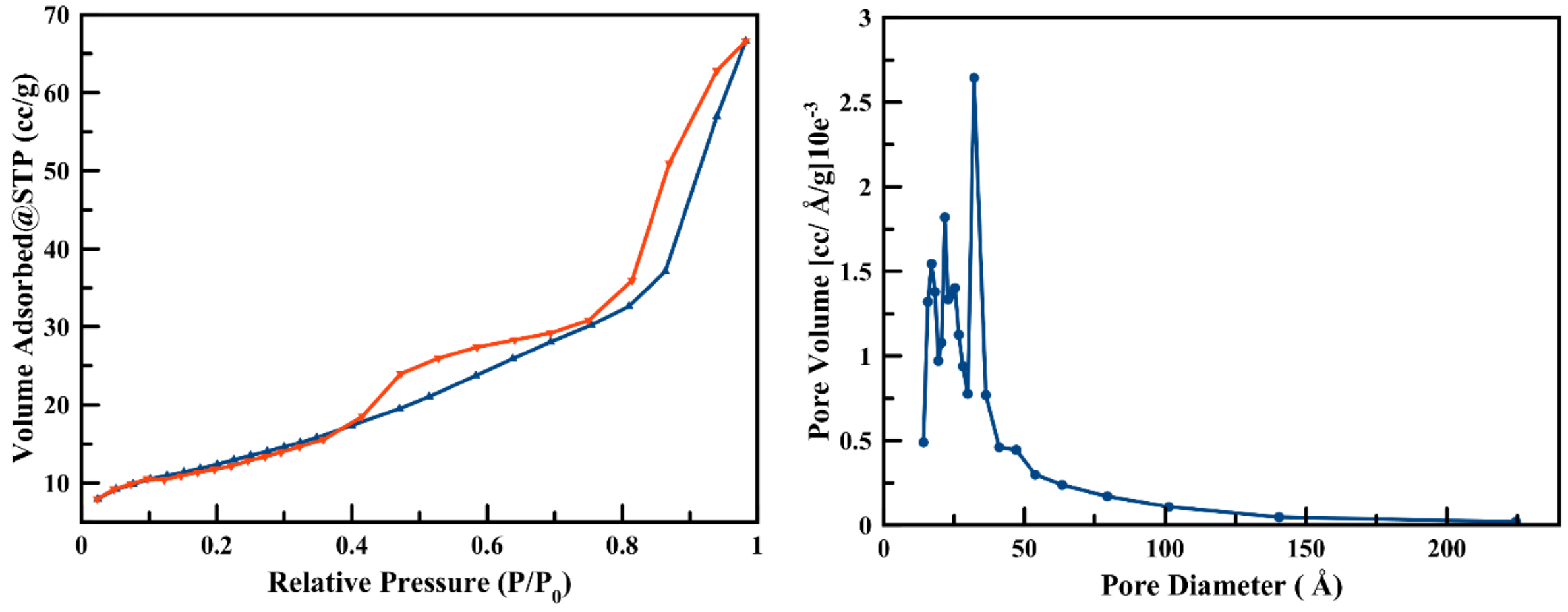

Fig. 2 (a) $N_2$ adsorption/desorption isotherms and (b) pore size distribution of $SnO_2$/BN composite

Addition of $SnO_2$ into boron nitride nanosheets leads to the separation of stacked BN nanosheets. This separation of sheets with addition of $SnO_2$ provided high surface area. Accumulation of $SnO_2$ in boron nitride nanosheets is shown in Fig. 3.

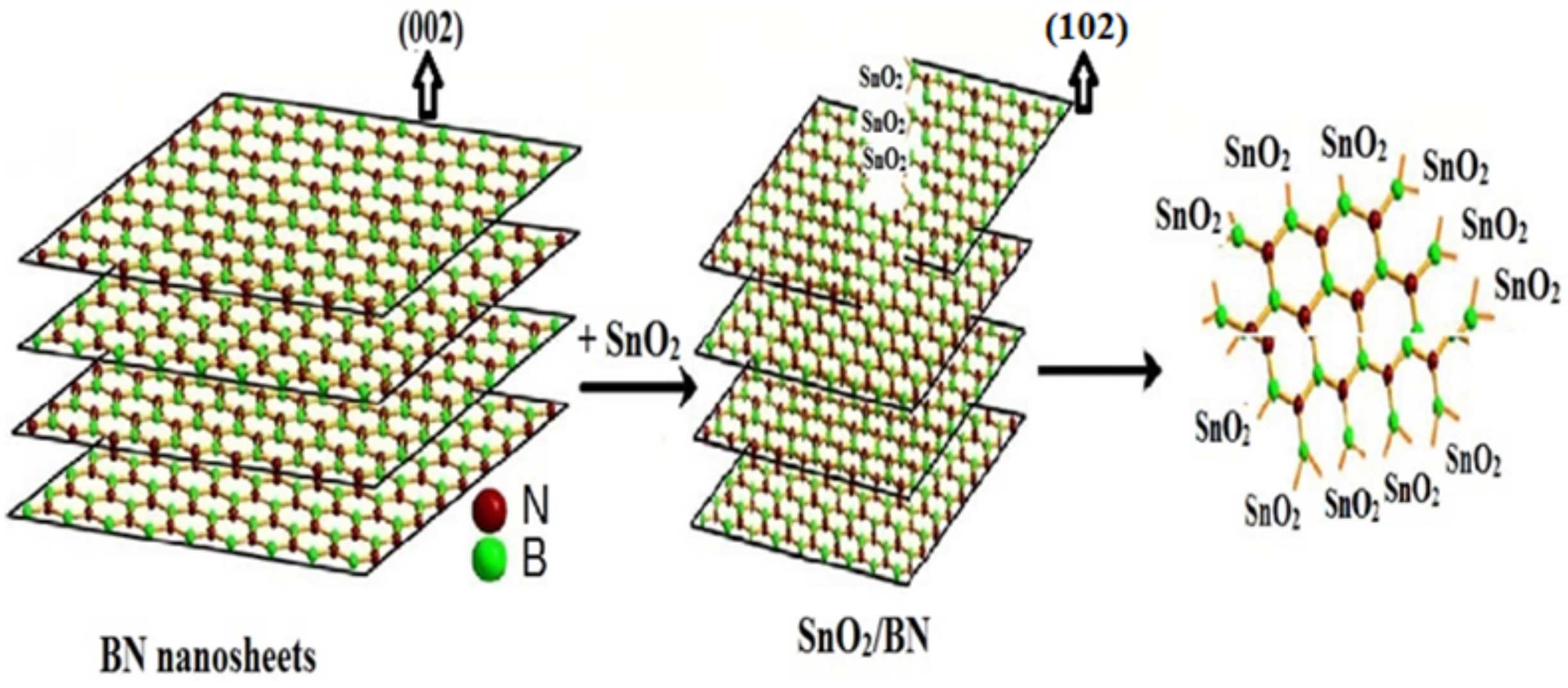

Fig. 3 Pictorial representation of accumulation of $SnO_2$ in boron nitride nanosheets

The texture coefficient provides the information regarding the growth direction of the structures. Texture coefficient is calculated using the relation given below [31, 32]

$$P(h_i k_i l_i) = \frac{I(h_i k_i l_i)}{I_0(h_i k_i l_i)} \left[ \frac{1}{n} \sum_{i=1}^{n} \frac{I(h_i k_i l_i)}{I_0(h_i k_i l_i)} \right]^{-1} \quad (2)$$

Where P (hkl) is texture coefficient of the plane specified by miller indices, I (hkl) and $I_0$ (hkl) are the specimen and standard intensities respectively for a given peak and n is the number of different peaks. The calculated values of texture coefficient are given in table 1. (102) plane with least density is able to provide more active sites which is useful in photocatalytic process.

**Table 1:** Texture coefficient analysis of BN and composite BN/$SnO_2$

| **Sample** | **Texture Coefficient** | | |
|---|---|---|---|
| | **(002)** | **(100)** | **(102)** |
| BN | 1.9999 | 0.1148 | 1.261 |
| BN/$SnO_2$ | 0.415 | 0.139 | 2.449 |

To study the morphology of samples, FESEM images were taken for the boron nitride, $SnO_2$ nanoparticles, and BN/$SnO_2$ nanocomposite. Fig. 4 (a) shows clear two-dimensional (2D) layered structure of boron nitride. Fig. 4 (b) shows the spherical morphology of $SnO_2$ nanoparticles. Fig. 4 (c) shows that 2D boron nitride sheets are decorated by numerous particles. Fig. 5 shows the elemental mapping images of BN/$SnO_2$ composite. These images clearly showed B, N, Sn and O elements evenly spread over entire area of the sample confirming good dispersion of BN nanosheets in the $SnO_2$ nanoparticles.

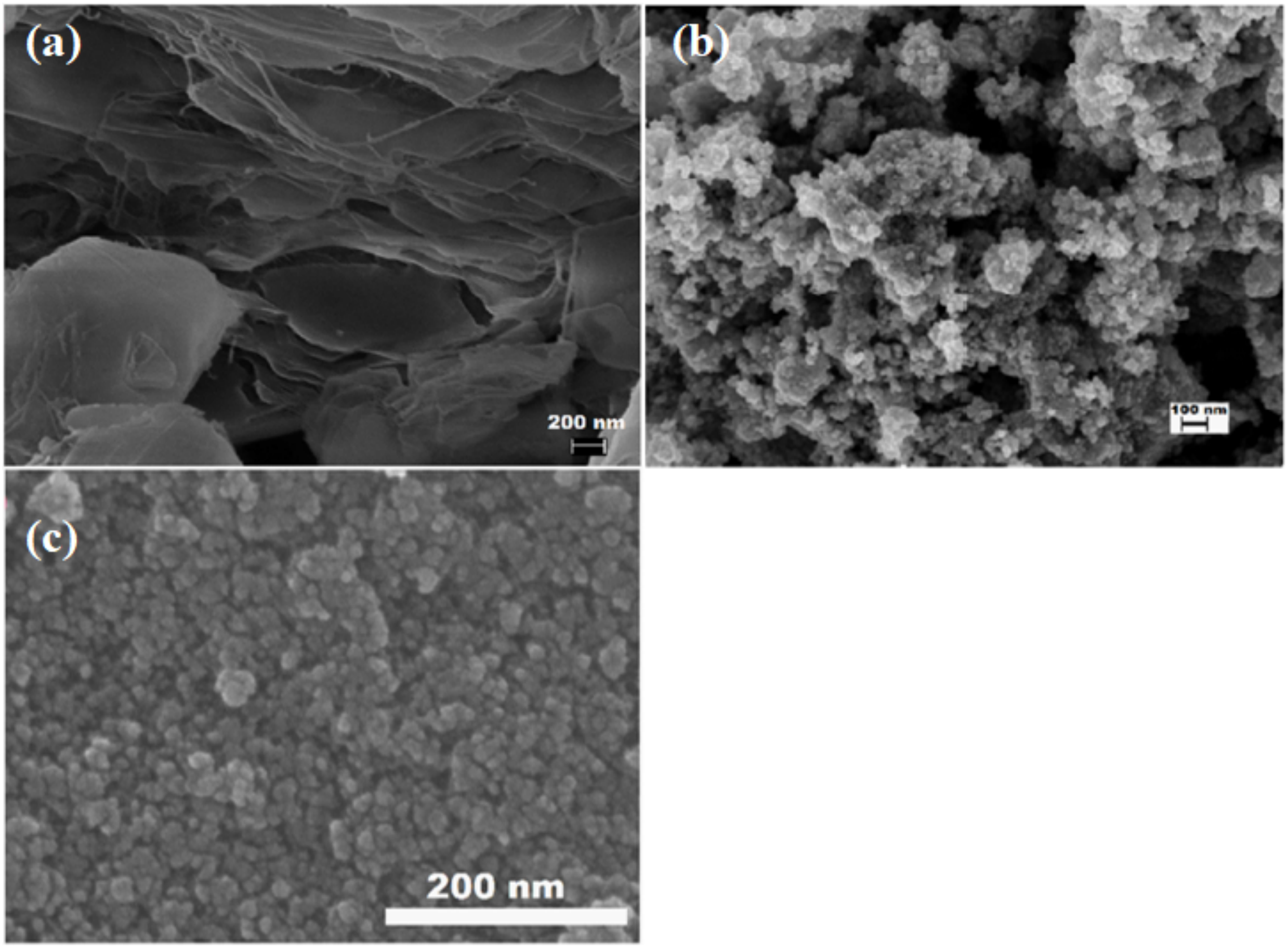

Fig. 4 FESEM images of (a) boron nitride, (b) $SnO_2$, (c) $BN/SnO_2$ composite respectively

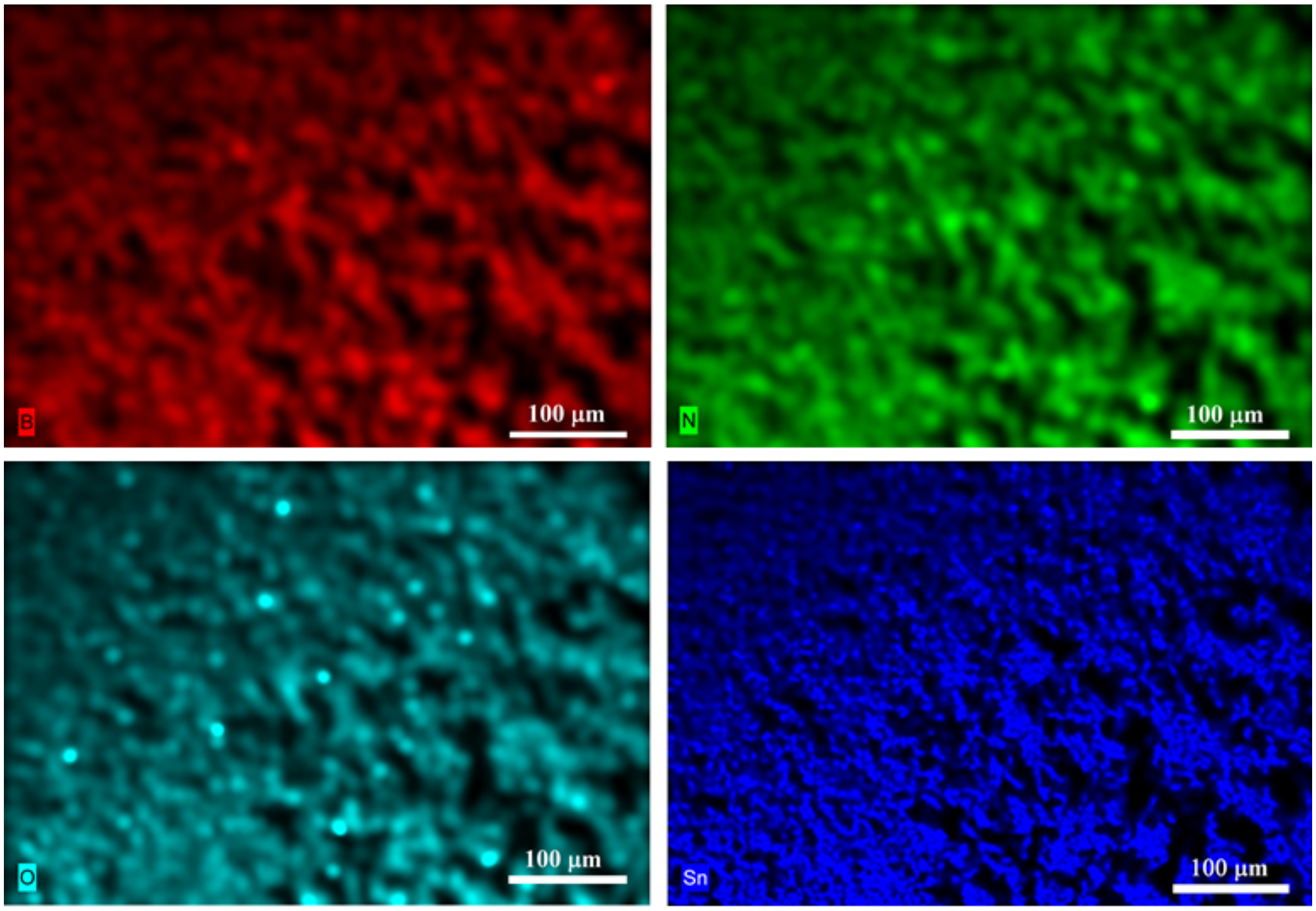

Fig. 5 Element mapping images of the BN/$SnO_2$ composite

Furthermore, the elemental analysis was performed using XPS. Fig. 6 (a) shows the XPS survey scan of BN/$SnO_2$ composite containing B 1s, N 1s, Sn $3d_{5/2}$, Sn $3d_{3/2}$, O 1s, Sn $3p_{3/2}$ and Sn $3p_{1/2}$ peaks. A small peak at 198.6 shown in Fig. 6 (a) revealed the presence of boron in composite sample. Peak-to-peak separation between Sn $3d_{5/2}$ and Sn $3d_{3/2}$ binding energy is in agreement with the reported value (8.5 eV) [33] of $SnO_{2-x}$ as shown in Fig. 6 (c). Similarly, the Sn $3d_{5/2}$ peak revealed the adsorbed species of $O^{-}$ (O–$Sn^{4+}$) and $O^{2-}$ (O–$Sn^{2+}$) on $SnO_2$ surface. The peak-to-peak separation increases from 8.5 eV to 8.58 eV, indicating the binding between BN and $SnO_2$ in the composite [34,35]. Clearly, the formation of bonds Sn-B-O bonds (Fig. 6 (d)), confirmed the strong interactions between the $SnO_2$ nanoparticles and BN surface. These strong interactions are the key factors for this synergistic effect [36].

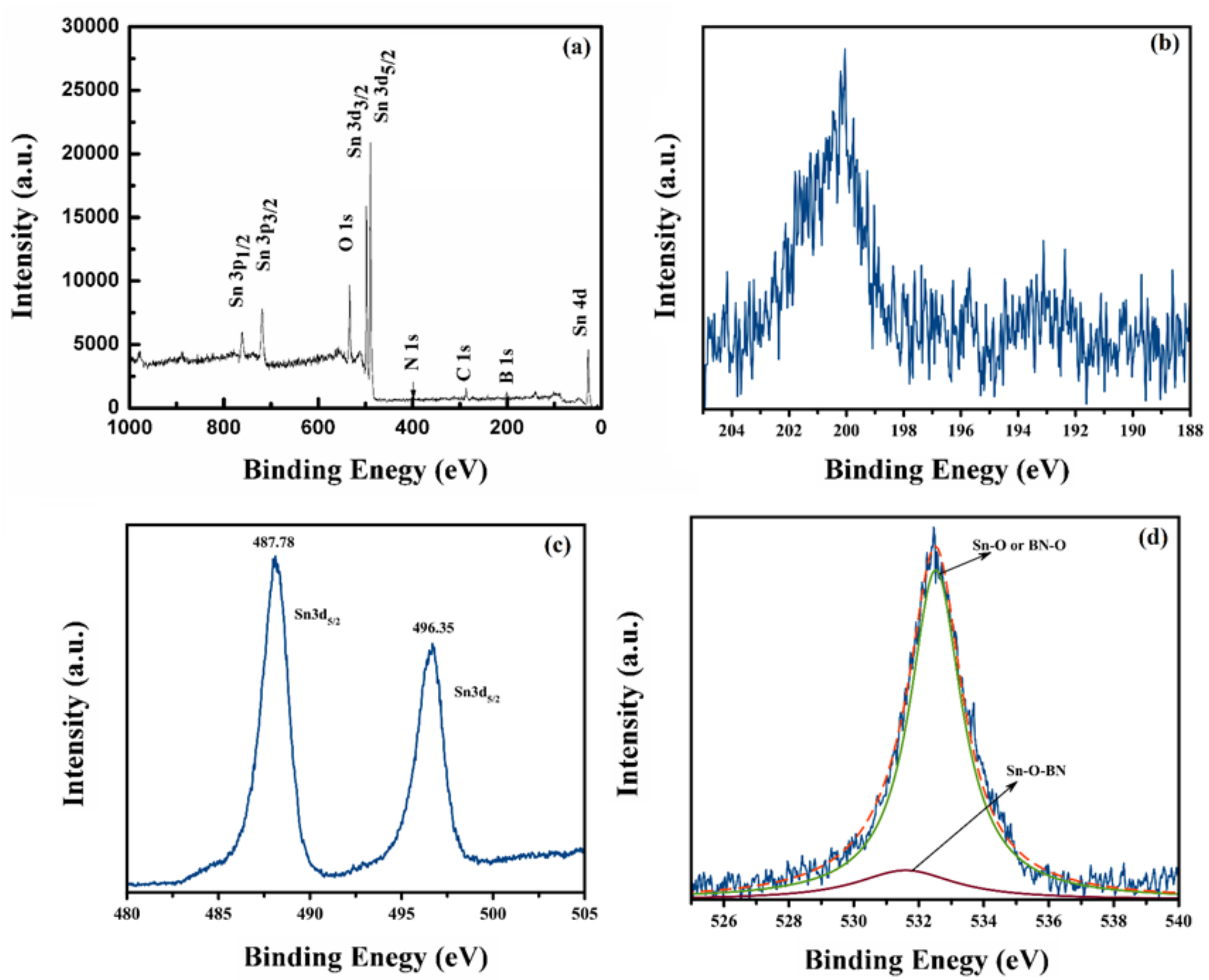

Fig. 6 XPS spectra of $SnO_2$/ BN composite, (a) survey (b) B 1s (c) O 1s and (d) Sn 3d respectively

UV-Vis absorbance spectrum of BN, $SnO_2$ and BN/$SnO_2$ nanocomposite is shown in Fig. 7. It can be perceived that the absorption edge of BN/$SnO_2$ is red-shifted to 427 nm compared to BN and $SnO_2$, indicating photo-response to visible light irradiation. The direct band gap energies of BN/$SnO_2$ photocatalysts were measured to be approximately 2.9 eV. The reduction of band gap energy may be attributed to the flawed crystallization of $SnO_2$ nanoparticles which supports the properties [22,37]. Growth of active $SnO_2$ on the surface of few-layered BN nanosheets not only serve as a stable support to prevent aggregation of $SnO_2$ nanoparticles but also enrich the dye molecules on the surface of support materials, both of which benefit the photocatalysis.

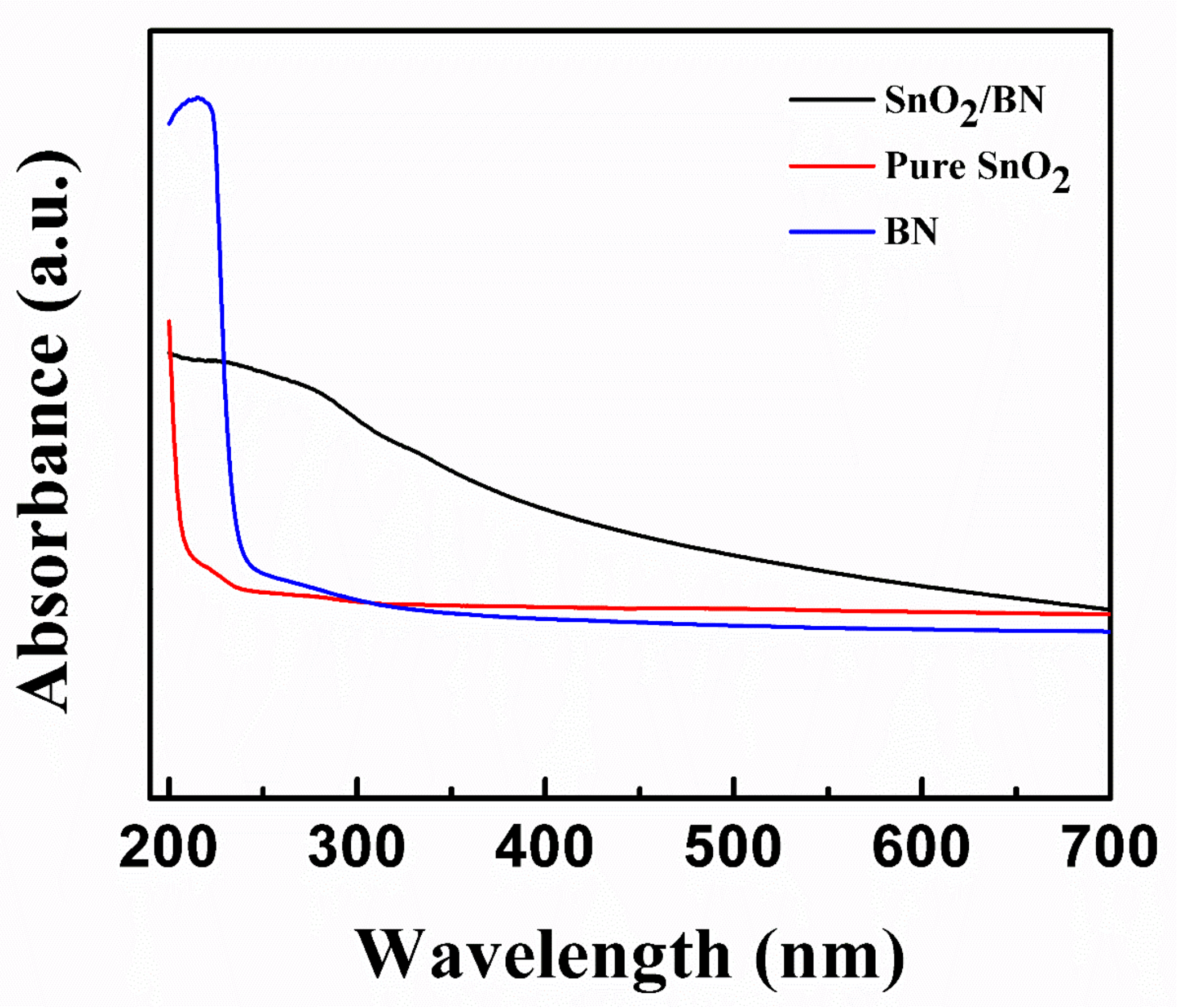

Fig. 7 UV-visible spectrum of nanostructured BN, $SnO_2$ and BN/$SnO_2$

Photocatalytic activity of BN, $SnO_2$ and BN/$SnO_2$ composite was determined by degradation of MO dye in aqueous solution. Fig. 8 (a) shows that there was no degradation in MO dye when only BN was used as a photocatalyst. When $SnO_2$ was used as a photocatalyst, the intensity of maximum adsorption peak positioned near 464 nm gradually decreased as shown in Fig. 8 (b) which shows the degradation of MO dye. For $SnO_2$, the degradation efficiency was calculated as ~42% in 40 minutes. For BN/$SnO_2$ composite the photocatalytic activity is much higher than that of $SnO_2$. During photodegradation of MO by BN/$SnO_2$ photocatalyst, the spectral absorbance degrades about 92% (Fig. 8 (c) and (d)) after only 7 minutes.

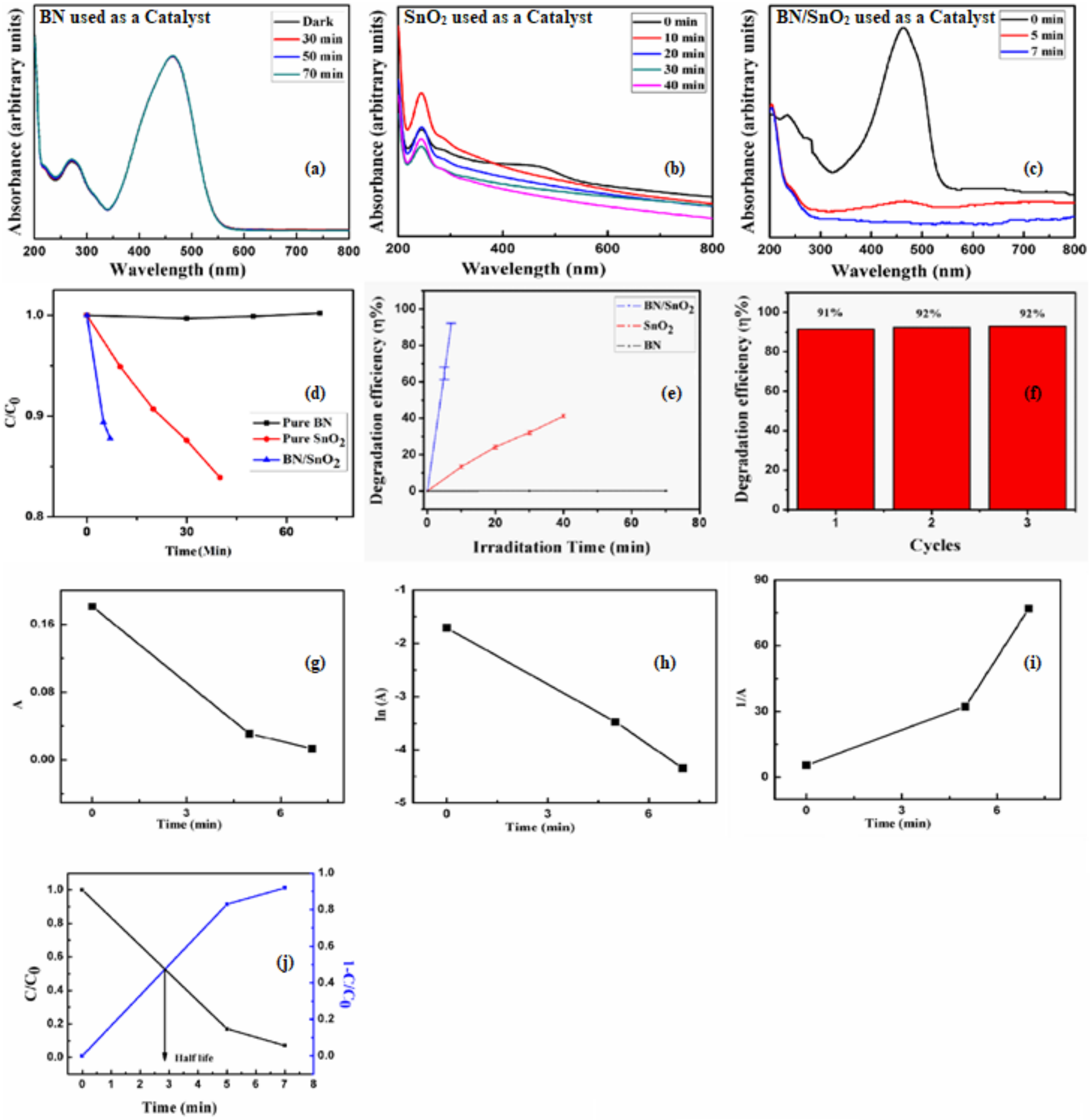

Fig. 8 UV-visible spectral changes MO using different catalysts (a) BN (b) $SnO_2$ (c) $SnO_2$/ BN (d) Concentration vs irradiation time of MO with nanostructured BN, $SnO_2$ and BN/$SnO_2$ nanocomposite (e) Degradation efficiency of different catalysts towards MO dye (f) Recycled photoactivity testing of BN/$SnO_2$ for degradation of MO (g-i) Pseudo-first-order reaction kinetics for $SnO_2$/ BN composite (j) $C/C_0$ curve and degradation efficiency ($1-C/C_0$) for BN/$SnO_2$

From the graph, it can be demonstrated that the MO dye with BN shows stability when irradiated with visible light. On the other hands, in the presence of BN/$SnO_2$ catalyst, the degradation rate remarkably enhanced as shown in Fig. 8 (e) Hence, BN/$SnO_2$ can be considered as an excellent photocatalyst. The stability of photocatalysts is important in large-scale processes for photocatalytic property. The stability results of BN/$SnO_2$ photocatalysts are shown in Fig. 8 (f). Fig. 8 (g-i) gives a plot between irradiation time and ln (A). The plot gives linear relationship which indicated that MO degradation follows a pseudo-first-order reaction kinetics described by equation [38,39],

$$\ln\frac{-d[A]}{dt} = k[A] \quad (3)$$

here [A] correspond to the initial concentration of MO solution at time 't', while 'k' is the photodegradation rate constant. The rate constants for the photo-degradation reactions are given in table 2. The half-life of the dye which is the time required for the MO concentration to decrease by half, was determined from the intersection point of curves between changes in MO concentration ($C/C_0$) and degradation efficiency ($1-C/C_0$). When BN/$SnO_2$ composite used as a catalysis the photodegradation efficiency was increased and half-life decreased on the basis of curves (Fig. 8 (j) and table 2).

**Table 2:** Photodegradation data of MO with BN, Pure $SnO_2$ and Composite BN/$SnO_2$ under visible light irradiation

| Material + Dye | Half-life (min) | Degradation efficiency (%) | Rate constant ($s^{-1}$) |
|---|---|---|---|
| BN + MO | - | No degradation or very low | - |
| $SnO_2$ + MO | - | 42 | - |
| BN/$SnO_2$ + MO | ~3 | 92 | $6.3 \times 10^{-3} s^{-1}$ |

Photocatalysis is usually the oxidation of organic compounds in which the absorption of light energy is the driving reaction force. The excited electrons decompose the organic molecule into the free radicals which are extremely reactive and also the excited point of the molecule is attacked by oxygen molecules, which then oxidise the organic compound [40].

In the present case of the BN/$SnO_2$ photocatalysts degrading MO dye, the possible photocatalytic process is illustrated as follows in equations (4−10):

$$MO + hv \rightarrow MO^* \quad (4)$$

$$MO^* + SnO_2 \rightarrow MO\,(h^+) + SnO_2(e^-) \quad (5)$$

$$MO^* + BN \rightarrow MO\,(h^+) + BN(e^-) \quad (6)$$

$$BN(e^-) + SnO_2 \rightarrow BN + SnO_2(e^-) \quad (7)$$

$$SnO_2(e^-) + O_2 \rightarrow SnO_2 + O_2^{-\bullet} \quad (8)$$

$$O_2^{-\bullet} + H_2O \rightarrow OH^\bullet \quad (9)$$

$$Dye + OH^\bullet \rightarrow Degraded\ products \quad (10)$$

In the study of BN/$SnO_2$ nanocomposite, due to the photoactivation the photoinduced electron $e^-$ are transferred into nanosheets of h-BN and combined with hole $h^+$ from donor species ($H_2O$) to form $OH^\bullet$ which oxidize the dyes; the $e^-$ reduce the electron acceptor ($O_2$) to form $O_2^{-\bullet}$ which participated in cleavage of the organic molecule (dismutation to $H_2O_2$). The recombination of $e^-$ and $h^+$ is inhibited as shown in Fig. 9 (a).

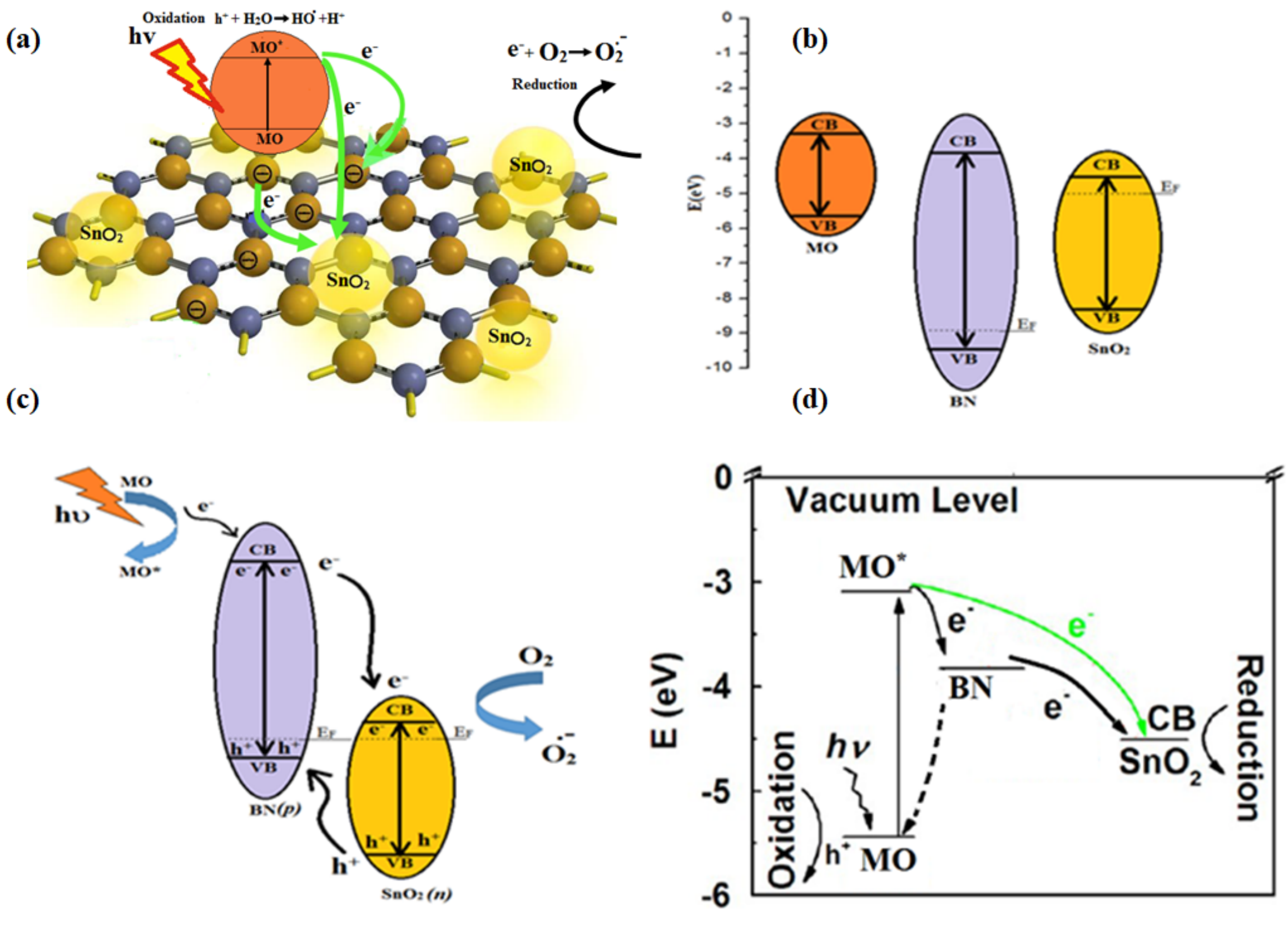

Fig. 9 (a) Inhibition of electron-hole recombination (b) Schematic band diagrams of MO, BN and $SnO_2$ showing conduction band and valance band (c) Fermi level equilibrium of BN and $SnO_2$ and possible transfer of electron-hole pairs, and (d) Schematic illustration of photosensitized degradation of the MO dyes Over the BN/$SnO_2$ photocatalyst under visible irradiation

The production of electrons and holes in pure BN and $SnO_2$ under visible light is expectable due to the presence of defect states in the band gap. The schematic band diagrams of pure BN and $SnO_2$ are shown in Fig. 9 (b). $SnO_2$ is a n-type (presence of donor states) semiconductor and BN is a p-type (presence of acceptor states) semiconductor, so the electron concentration in $SnO_2$ is much higher than that of BN. Furthermore, the Fermi energy level of the n-type $SnO_2$ is close to the conduction band (CB), which is close to the valence band for p-type BN [41]. Furthermore, the CB potential of n-type $SnO_2$ was more positive than that of p-type BN, and the VB potential of p-type BN was more negative than that of n-type $SnO_2$ [42]. The visible light irradiates MO to MO*. Then the excited MO* can inject photogenerated electrons into BN layer and $SnO_2$ using a sensitization process as shown in Fig. 9 (c). The excited MO* also creates photogenerated hole ($MO^+$). The hindrance of MO degradation would occur by recombination of injected electron ($e^-$) and photogenerated hole ($MO^+$). Fortuitously, due to $SnO_2$ into BN layers, BN acts as an electron mediator which helps in migration of electron from MO* to conduction band of $SnO_2$ (eqn. 6, 7) shown in Fig. 9 (d) [43-45]. BN causes main hindrance for recombination of electron and holes in MO and facilitate migration of electron transfer i.e. from excited MO* and excited BN to CB of $SnO_2$. The reduction and oxidation reactions occurred in conduction band of $SnO_2$ and ground state of MO. Photo induced catalysis of these systems proceed via $OH^\bullet$ radical formation in reaction medium.

To recognize the presence of $OH^\bullet$ radical, we have done a typical experiment using terephthalic acid (TA) as a probe which reacts with reactive $OH^\bullet$ radical and generate 2-hydroxyterethalic acid (TAOH), a luminescent active species. The luminescence intensity of TAOH is proportional to the amount of $OH^\bullet$ radical produced under visible light illumination. Upon excitation at 315 nm and for ensuring the visible light irradiation, an ultraviolet cutoff filter ($\lambda$=400 nm) was used. The maximum intensity at 425 nm was measured at 10-minute time interval. It is seen from Fig. 10 (a) that luminescence intensity gradually increased with increasing irradiation time. This consequence indicates that $OH^\bullet$ radical definitely produced at the time of photocatalysis. Fig. 10 (b) shows that the luminescence intensity at 425 nm linearly increases with irradiation time.

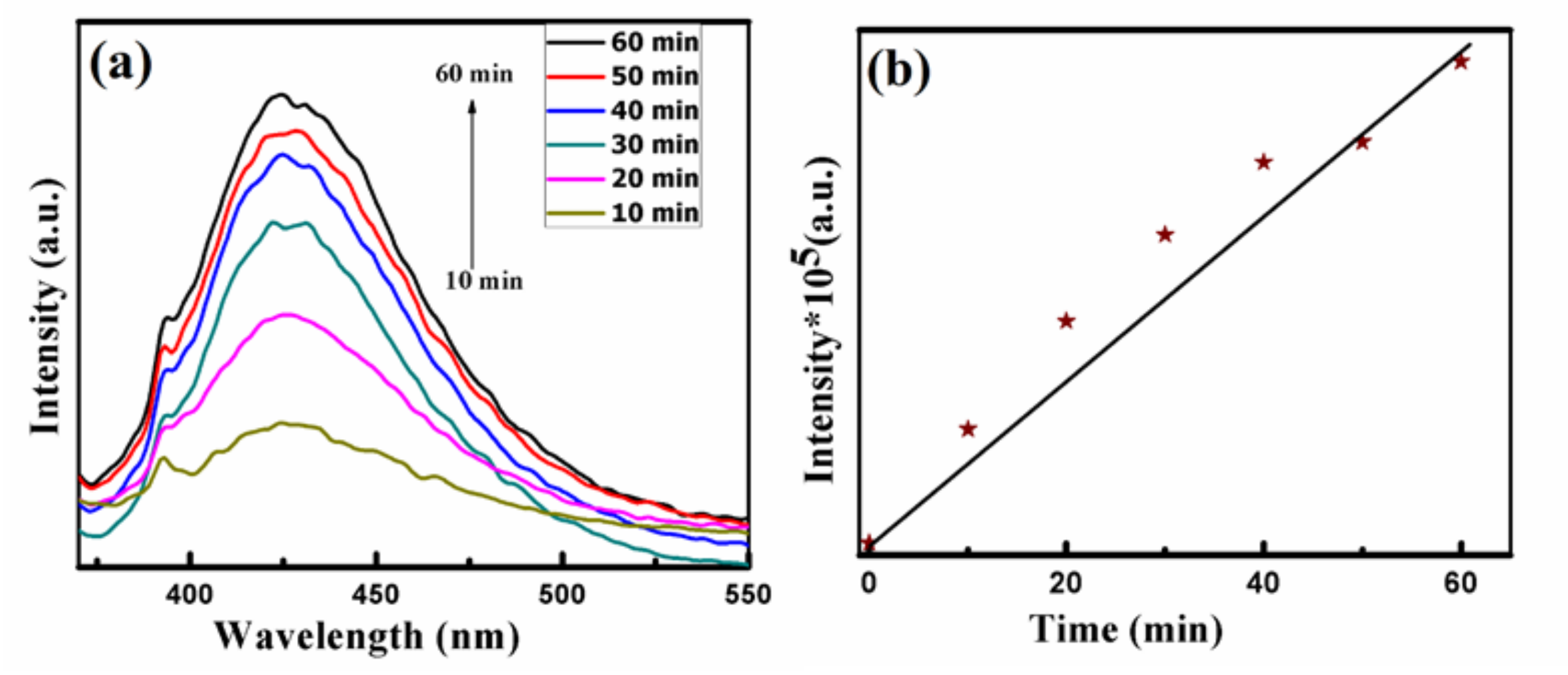

Fig. 10 (a) Intensity variation with time for TA solution in the presence of $SnO_2$/BN composite under light irradiation, (b) Intensity at 425 nm against illumination time for TAOH

There are several factors which plays prominent role in the enhanced photocatalytic activity of BN/$SnO_2$. XRD analysis shows that small sized BN/$SnO_2$ composite increases the particles surface areas which facilitate more photocatalytic active sites. Also, the band gap of BN/$SnO_2$ lies in the visible light region which converts the excitation light to chemical energy to accelerate photocatalytic oxidation. BN/$SnO_2$ exhibits strong adsorption ability for MO due to its layered structure, hydrophobic property and high surface area, thus MO will concentrate around the loaded $SnO_2$ nanoparticles. The good contact between the MO and $SnO_2$ are beneficial to enhance the photocatalytic efficiency of BN/$SnO_2$. Moreover, layered BN in contact with $SnO_2$ also suppress the recombination of photogenerated electrons and holes at BN/$SnO_2$ interfaces, allowing both of them to participate in the photocatalytic reaction.

It is well known fact that MO is a self-degraded dye, so we have decided to use some other colorless pollutant for degradation process i.e. salicylic acid. Photocatalytic activity of BN/$SnO_2$ nanocomposite for salicylic acid is shown in Fig. 11 (a). BN/$SnO_2$ composite degraded salicylic acid about 82% in 40 min as shown in Fig. 11 (b). This experiment confirms that BN/$SnO_2$ nanocomposite is very effective in degradation of self-degradation as well as colorless pollutant.

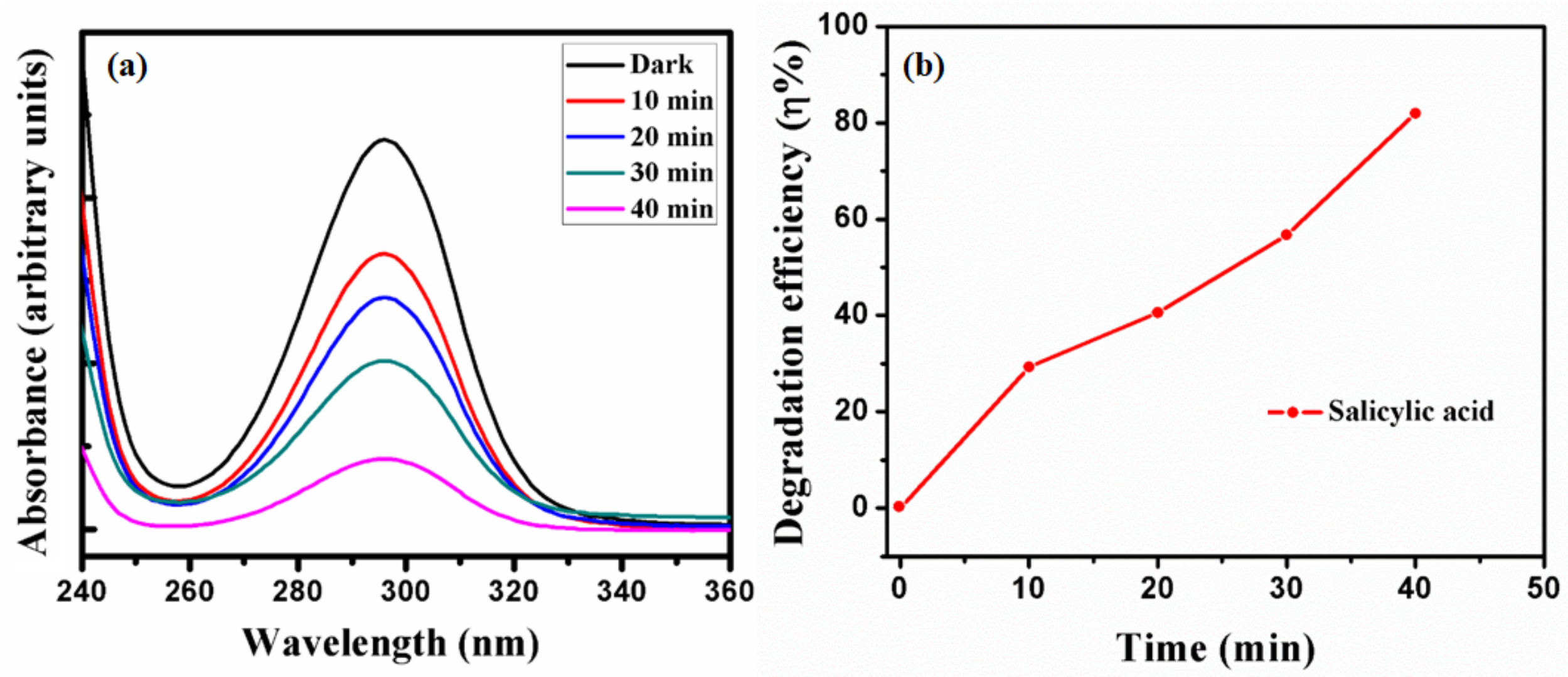

Fig. 11 (a) UV-visible spectral changes of photodegradation of salicylic acid using $SnO_2$/ BN as Catalyst (b) Degradation efficiency of BN/$SnO_2$ catalysts towards salicylic acid

## 4. Conclusion

BN/$SnO_2$ composite has been synthesized by hydrothermal cum-wet chemical method. XRD analysis confirmed the formation of BN and $SnO_2$ phases in the prepared composite. Preferential growth of the synthesized material leads to the formation large number of surface-active sites which are responsible for high photocatalytic properties. Results revealed that the synthesized composite exhibit enhanced photocatalytic degradation for methylene orange organic dye under sunlight irradiation i.e. BN/$SnO_2$ catalyst degraded MO dye under 7 minutes. Due to its high photocatalytic activity (92 % degradation) this composite material shows good capability for use in environmental as well as industrial applications. BN/$SnO_2$ composite could be used in the degradation of other self- photosensitized dyes under sunlight irradiation.

## 5. Acknowledgements

This work was funded by Department of Science and Technology under Science and Engineering Research Board (SERB) project no. EMR/2016/002815.

## References

[1] L. Jing, W. Zhou, G. Tian, H. Fu, Chem. Soc. Rev. 42 (2013) 9509–9549.

[2] L.Yang, C. Xu, F. Wan, H. He, H. Gu, J. Xiong, G. Wang, Materials Research Bulletin. 112 (2019) 154-158.

[3] P. Zhou, J. Yu, M. Jaroniec, Adv. Mater. 26 (2014) 4920–4935.

[4] X. Zhou,Y. Mu, S. Zhang, L. Gao, H. Chen, J. Mu, X. Z, M. Zhang, W. Liu, Materials Research Bulletin. 111 (2019) 118-125.

[5] A.B. Djurišić, Y.H. Leung, A.M. Ching Ng, Mater. Horizons. 1 (2014) 400.

[6] S.A. Mahmoud, O.A. Fouad, Sol. Energy Mater. Sol. Cells. 136 (2015) 38–43.

[7] S. Aghabeygi, Z. Sharifi, N. Molahasani, Dig. J. Nanomater. Biostructures. 12 (2017) 81-89.

[8] J. Jeon, K. Kon, T. Toyao, K. Shimizu and S. Furukawa, Chem. Sci. (2019) Advance Article.

[9] Z. Wang, C. Li, K. Domen, Chem. Soc. Rev. (2019) Advance Article.

[10] X. Li, J. Zhao, J. Yang, Scientific reports, 3 (2013) 1858.

[11] S.Z. Butler, S.M. Hollen, L. Cao, Y. Cui, J.A. Gupta, H.R. Gutiérrez, T.F. Heinz, S.S. Hong, J. Huang, A.F. Ismach, ACS Nano. 7 (2013) 2898–2926.

[12] D. Liu, M. Zhang, W. Xie, L. Sun, Y. Chen, W. Lei, Appl. Catal. B Environ. 207 (2017) 72–78.

[13] R. Zhang, J. Wang, P. Han, Alloys Compd. 637 (2015) 483–488.

[14] Y. He, L. Zhang, M. Fan, X. Wang, M.L. Walbridge, Q. Nong, Y. Wu, L. Zhao, Sol. Energy Mater. Sol. Cells. 137 (2015) 175–184.

[15] G. Postole, A. Gervasini, M. Caldararu, B. Bonnetot, A. Auroux, Appl. Catal. A Gen. 325 (2007) 227–236.

[16] W. Sun, Y. Meng, Q. Fu, F. Wang, G. Wang, W. Gao, X. Huang, F. Lu, ACS Appl. Mater. Interfaces. 15 (2016) 9881–9888.

[17] X. Hu, H. Zhao, J. Tian, J. Gao, Y. Li, H. Cui, Sol. Energy Mater. Sol. Cells. 172 (2017) 108–116.

[18] Y. Song, H. Xu, C. Wang, J. Chen, J. Yan, Y. Xu, Y. Li, C. Liu, H. Li, Y. Lei, RSC Adv. 4 (2014) 56853–56862.

[19] C. Tang, J. Li, Y. Bando, C. Zhi, D. Golberg, Chem. Asian J. 5 (2010) 1220–1224.

[20] S. Meng, X. Ye, X. Ning, M. Xie, X. Fu, S. Chen, Appl. Catal. B Environ. 182 (2016) 356–368.

[21] X. Fu, Y. Hu, T. Zhang, S. Chen, Appl. Surf. Sci. 280 (2013) 828–835.

[22] H. Wang, F. Sun, Y. Zhang, L. Li, H. Chen, Q. Wu, C.Y. Jimmy, J. Mater. Chem. 20 (2010) 5641–5645.

[23] G. Wang, W. Lu, J. Li, J. Choi, Y. Jeong, S.-Y. Choi, J.-B. Park, M.K. Ryu, K. Lee, Small. 2 (2006) 1436–1439.

[24] M. Wang, M. Li, L. Xu, L. Wang, Z. Ju, G. Li, Y. Qian, Catal. Sci. Technol. 1 (2011) 1159–1165.

[25] B. Singh, P. Singh, M. Kumar, A. Thakur, A. Kumar, AIP Conf. Proc., 2015: p. 80024.

[26] B. Singh, G. Kaur, P. Singh, K. Singh, B. Kumar, A. Vij, M. Kumar, R. Bala, R. Meena, A. Singh, A. Thakur, A. Kumar, Sci. Rep. 6 (2016) 35535.

[27] B. Singh, G. Kaur, P. Singh, K. Singh, J. Sharma, M. Kumar, R. Bala, R. Meena, S.K. Sharma, A. Kumar, New J. Chem. 41 (2017) 11640–11646. doi:10.1039/C7NJ02509B.

[28] M. Montazerozohori, M. Nasr-Esfahani, S. Joohari, Environ. Prot. Eng. 38 (2012) 45–55.

[29] G. Kaur, B. Singh, P. Singh, M. Kaur, K. K. Buttar, K. Singh, A. Thakur, R. Bala, M. Kumar, A. Kumar, RSC Adv. 6 (2016) 99120–99128

[30] S. Khanchandani, S. Kundu, A. Patra, A.K. Ganguli, J. Phys. Chem. C. 117 (2013) 5558–5567.

[31] M. Kumar, A. Kumar, A.C. AbhyankarACS Appl. Mater. Interfaces. 7 (2015) 3571–3580.

[32] V. Kumar, K. Singh, M. Jain, Manju, A. Kumar, J. Sharma, A. Vij, A. Thakur, Applied Surface Science 444 (2018) 552–558.

[33] L. Chen, L. Xie, M. Wang, X. Ge, J. Mater. Chem. A. 3 (2015) 2991–2998.

[34] H. Seema, K.C. Kemp, V. Chandra, K.S. Kim, Nanotechnology. 23 (2012) 355705.

[35] A. Kumar, L. Rout, R.S. Dhaka, S.L. Samal, P. Dash, RSC Adv. 5 (2015) 39193–39204.

[36] R. Tian, Y. Zhang, Z. Chen, H. Duan, B. Xu, Y. Guo, H. Kang, H. Li, H. Liu, Sci. Rep. 6 (2016) 19195.

[37] L. Zhu, M. Hong, G. W. Ho. Scientific Reports. 5 (2015) 11609.

[38] G. Kaur, B. Singh, P. Singh, K. Singh, A. Thakur, M. Kumar, R. Bala, A. Kumar, Chemistry Select. 2 (2017) 2166–2173.

[39] Y. Wu, S. Zeng, F. Wang, M. Megharaj, R. Naidu, Z. Chen, Separation and Purification Technology. 154 (2015) 161–167.

[40] V. Štengl, J. Henych, M. Slušná, J. Nanomater. 2016 (2016) Article ID 4580516.

[41] Y. Xu, M.A.A. Schoonen, Am. Mineral. 85 (2000) 543–556.

[42] J. Cao, C. Zhou, H. Lin, B. Xu, S. Chen, Appl. Surf. Sci. 284 (2013) 263–269.

[43] F. Schulz, R. Drost, S.K. Hämäläinen, T. Demonchaux, A.P. Seitsonen, P. Liljeroth, Phys. Rev. B. 89 (2014) 235429.

[44] J. Zhang, K.P. Loh, M. Deng, M.B. Sullivian, J. Zheng, P. Wu, J. Appl. Phys. 99 (2006) 104309.

[45] Q. Weng, Y. Ide, X. Wang, X. Wang, C. Zhang, X. Jiang, Y. Xue, P. Dai, K. Komaguchi, Y. Bando, D. Golberg, Nano Energy. 16 (2015) 19–27.